\documentclass[pre,twocolumn,superscriptaddress,showpacs]{revtex4}
\usepackage{graphicx}
\usepackage{psfrag}

\begin{document}

\title{Clustering as a measure of the local topology of networks}

\date{\today}

\author{Alexandre H. Abdo}
\homepage{http://cecm.usp.br/~eris/}
\email{abdo@member.fsf.org}
\affiliation{Instituto de F\'{i}sica, Universidade de S\~{a}o Paulo, CP 66318, 05315-970 S\~{a}o Paulo, SP, Brazil}
\author{A. P. S. de Moura}
\affiliation{College of Physical Sciences and Engineering, Univ. of Aberdeen, King's College, Aberdeen, AB24 3UE, UK}
\affiliation{Instituto de F\'{i}sica, Universidade de S\~{a}o Paulo, CP 66318, 05315-970 S\~{a}o Paulo, SP, Brazil}

\begin{abstract}
Usual formulations of the clustering coefficient can be shown to be insufficient in the task of describing the local topology of very simple networks. Motivated by this, we review some alternatives in order to present an extension, the {\em clustering profile}. We show, both conceptually and through applications to well studied networks, that this measure is a more complete and robust measure of clustering. It imposes stringent constraints on theoretical growth models, specially on aspects of the network structure that play a central role in dynamics on networks. In addition, we study how it provides a richer perspective of phenomena such as hierarchy, small-worlds and clusterization.
\end{abstract}


\maketitle

\section{Introduction}

Physicists have been greatly interested in studying the interplay between the topology and growth dynamics of complex networks, for they are a universal framework to understand various processes in previously distant areas ranging from cell metabolism to linguistics\cite{graphshandbook-editors, evolutionnetworks-dorogovtsev}. The unusual nature of their structure, represented by graphs\cite{graphs-diestel}, required the development of novel methods, and one of the core difficulties has been to understand what measurements are robust and indeed represent universal quantities by which we can compare and classify networks, or assert how effectively a network growth model reflects the network it tries to mimic.

Average distance, degree distribution and degree correlations\cite{networksreview-newman}, loopiness\cite{loopiness-bianconi}, motifs\cite{motifs-milo} and the clustering coefficient\cite{smallworld-watts} are some quantities which have established themselves as useful. But, among them, only the latter gives any description of the local topology --- how the network is organized close to some vertex --- measuring how many of its neighbors are also neighbors among themselves. It has also been established that local structure is not only an important topological characteristic, but also a main concern when studying searchability\cite{identitysearch-watts, freenet-zhang} and dynamics\cite{epidemicstructured-eguiluz, clusteringdisease-cross} on networks.

Still, important as it stands, the usual clustering coefficient has serious limitations, and is commonly replaced by \emph{ad hoc} definitions. We first focus on these shortcomings, as they'll let us understand why so many variations emerged, and by summarizing this variations propose a consistent and robust improvement, which should provide for a more complete and less specialized characterization of the local topology of networks.

We remark that our focus is not on clustering as transitivity\cite{serrano-generalclustering}, but as a general measure of small-scale structural organization.

\section{Background}

When first introduced\cite{smallworld-watts} to formalize the small-world phenomena, the clustering coefficient was defined as the fraction of connected pairs among the neighbors of a vertex, averaged over all vertices.

Following that, it was also defined by taking the fraction of connected pairs over all pairs of neighbors of all vertices\cite{generatingfunctions-newman}, since that frequently is more tractable. In this form, it was used to derive analytical expressions that distinguish social networks from other networks\cite{socialnetworks-newman}.

A first problem with the coefficient is that these two seemingly close definitions may yield very different results\cite{handbook-bollobas}, as can be seen from the fact that the individual ratio for vertices of high degree has greater impact on the second case.

Later, a better description of clustering was introduced by considering the coefficient as a function of vertex degree, and doing so revealed important structural features such as hierarchical organization\cite{hierarchical-barabasi}.

Another interesting consequence is that this also removes the ambiguity of the first two definitions: if averages are restricted to a set of vertices of same degree, averaging the coefficient over vertices or calculating the fraction of connections over all pairs of neighbors is the same thing.

Still, other definitions have become necessary, as in the case of bipartite networks used in the study of sexually transmitted diseases\cite{bipartite-herrmann}, where no odd cycles exist and thus, because connected neighbors of a vertex would form cycles of length $3$, the clustering coefficient is always zero, even though a bipartite network may have a very complex local structure.

The underlying issue becomes more evident when noting that both a square lattice and a single large cycle, which are locally very different networks, have their clustering coefficient equal to zero.

We take this as evidence that a general treatment of clustering cannot expect to rest on a simple scalar quantity; however, it should remain a local property that can be calculated for each vertex, and comparable among them. Thus the source of the problem seems to be the question asked, of ``how many of my neighbors are connected?'', instead of ``how closely related are my neighbors?'', which we think better comprehends the concept of clustering as a measure of local topology.

In order to implement this idea we turn back to the original definition of clustering for a vertex, the fraction of connected pairs among its neighbors, and notice that ``connected'' can stand for ``whose distance is $1$''. Then the usual clustering coefficient can be understood as the first term of a sequence, the term accounting for the fraction of pairs of neighbors whose distance is $1$. The second term would be the fraction of pairs of neighbors whose distance is $2$, and so on. Now, since those are all neighbors of the same vertex, they are always connected by a path of length $2$ going through that vertex, and so we must discard paths going through the vertex in question when calculating this distance between its neighbors.

These higher order clustering coefficients were first explored by A. Fronczak \emph{et al.}\cite{higherorder-fronczak} in order to show, in a stronger sense, that the model of preferential attachment\cite{scalefree-barabasi} is blind to clustering mechanisms.

\begin{figure}
\includegraphics[width=\columnwidth]{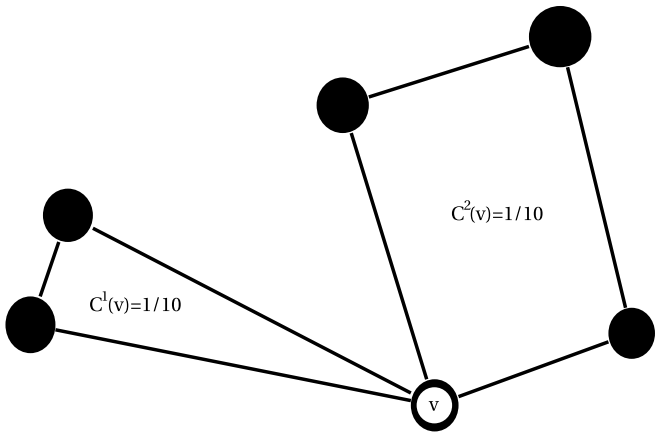}
\caption{\label{fig-cp}Higher orders of clustering, $C^{d}(v)$: the fraction of pairs of neighbors whose smallest cycle shared with the vertex has length $d$. In the example above, $v$ has $10$ pairs of neighbors and a non-zero clustering for $d=1,2$. $C^{1}(v)$ is, by definition, the usual clustering coefficient.}
\end{figure}

The meaning of this extended clustering can, perhaps, be more clearly understood in terms of cycles: the $n$th term of the sequence would correspond to the fraction of pairs of neighbors of a vertex whose smallest cycle shared with it has length $n+2$; as can be seen on FIG.~\ref{fig-cp}.

So, by combining these dependencies on degree and distance, we obtain a measure that is neither ambiguous nor specialized; which not only applies to all networks, but is a richer description of their local topology. For practical purposes, instead of referring to higher order clustering coefficients averaged as a function of degree, we name this quantity the clustering profile, and proceed to a formal definition.

\section{Clustering Profile}

In a network $G$ composed of vertices $V$ and edges $E$, so that $G=(V,E)$, we denote the clustering for a vertex $u\in V$ as $C^{d}(u)$, defined as the number of pairs of neighbors of $u$ whose distance in the induced network $G(V\setminus \{u\})$ is $d$, divided by the total number of pairs of neighbors of $u$. Thus

\begin{equation}
C^{d}(u) = \frac{ |\{ \{ v,w \}; v,w \in N(u) | d_{ G( V \setminus \{ u\} ) }(v,w) = d \} | }{ { |N(u)| \choose 2} } ,
\end{equation}

where $N(\cdot )$ is the set of neighbors of a vertex, the modulus $|\cdot |$ represents the cardinality (number of elements) of a set, and so $|N(u)|$ is the degree of $u$, also denoted $deg(u)$.

This leads to a generalized description of how the network is organized around that vertex, reflecting the contribution of more distant neighbors in higher terms, while still preserving the good property that, when summed over all terms, it ranges between $0$ and $1$.

We can then define the clustering profile for a network, being the average of $C^{d}(u)$ over all vertices of same degree $k$, and denote it

\begin{equation}
C^{d}_{k} = \frac{ \sum_{ \{u|deg(u)=k\} } C^{d}(u) }{ |\{u|deg(u)=k\}| } .
\end{equation}

It should be noticed that the usual clustering coefficient as a function of degree is simply $C^{1}_{k}$.

And although numerically calculating the clustering profile is a more expensive computer operation than calculating the usual coefficient, each step of the calculation is parallelizable, so even large networks can be treated with relatively small computer resources.

\section{Applications}

In order to illustrate the consequences of the clustering profile, we choose a well known network: the set of metabolic networks of bacteria first studied in references \cite{largescalemetabolic-barabasi,hierarchical-barabasi}, where a growth model\cite{hierarchical-barabasi} is provided to illustrate their hierarchical organization. Comparing the results for the real network and this hierarchical model will help us better understand the phenomena associated with this system. We first focus on the profile's absolute value, then its variation with distance, and finally its variation with degree, as each of these will have distinct implications.

\subsection{Small-worlds}

\begin{figure}
\psfrag{xlabel}[Bc][Bc]{$k$}
\psfrag{ylabel}[Bc][Bc]{$C^{d}_{k}$}
\psfrag{key0}[Br][Br][0.7]{$d\ 1$}
\psfrag{key1}[Br][Br][0.7]{$2$}
\psfrag{key2}[Br][Br][0.7]{$3$}
\psfrag{key3}[Br][Br][0.7]{$4$}
\psfrag{key4}[Br][Br][0.7]{$5$}
\psfrag{key5}[Br][Br][0.7]{$sum$}
\includegraphics[width=\columnwidth]{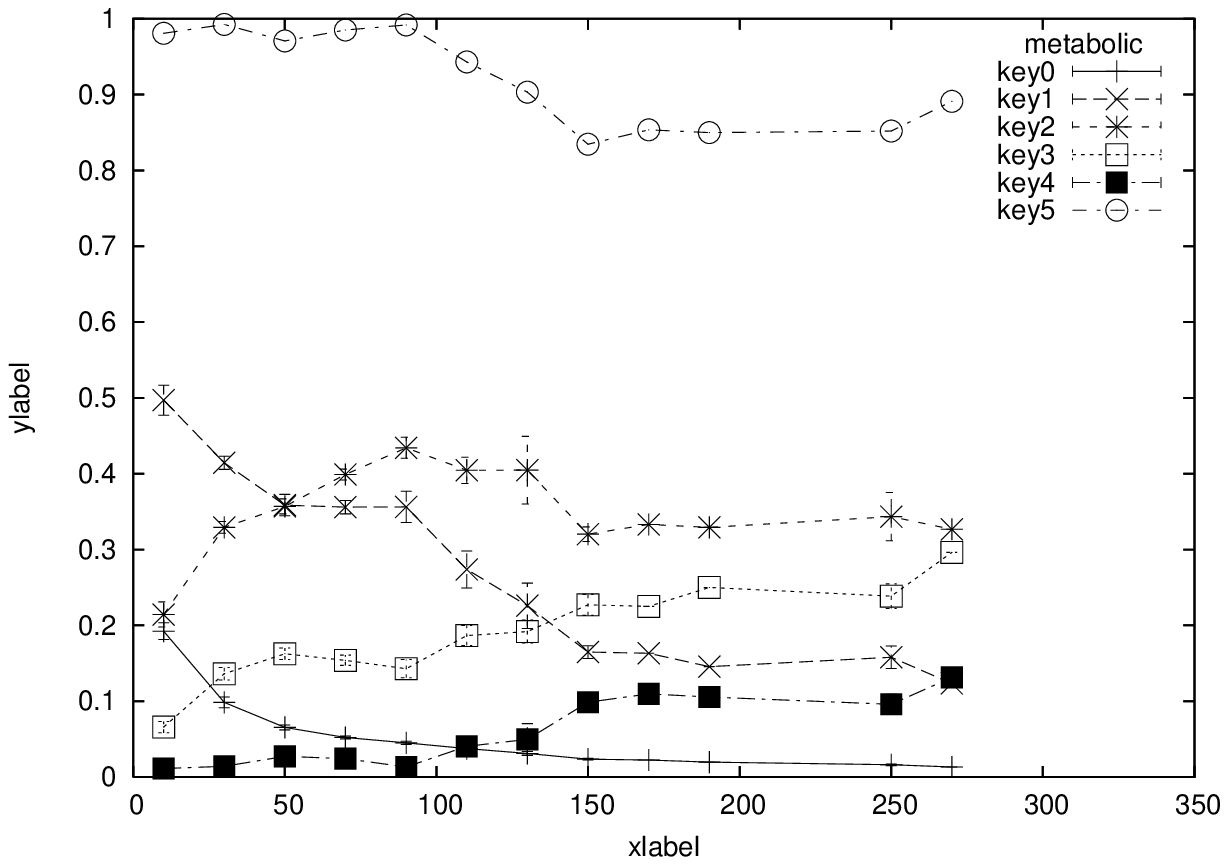}
\includegraphics[width=\columnwidth]{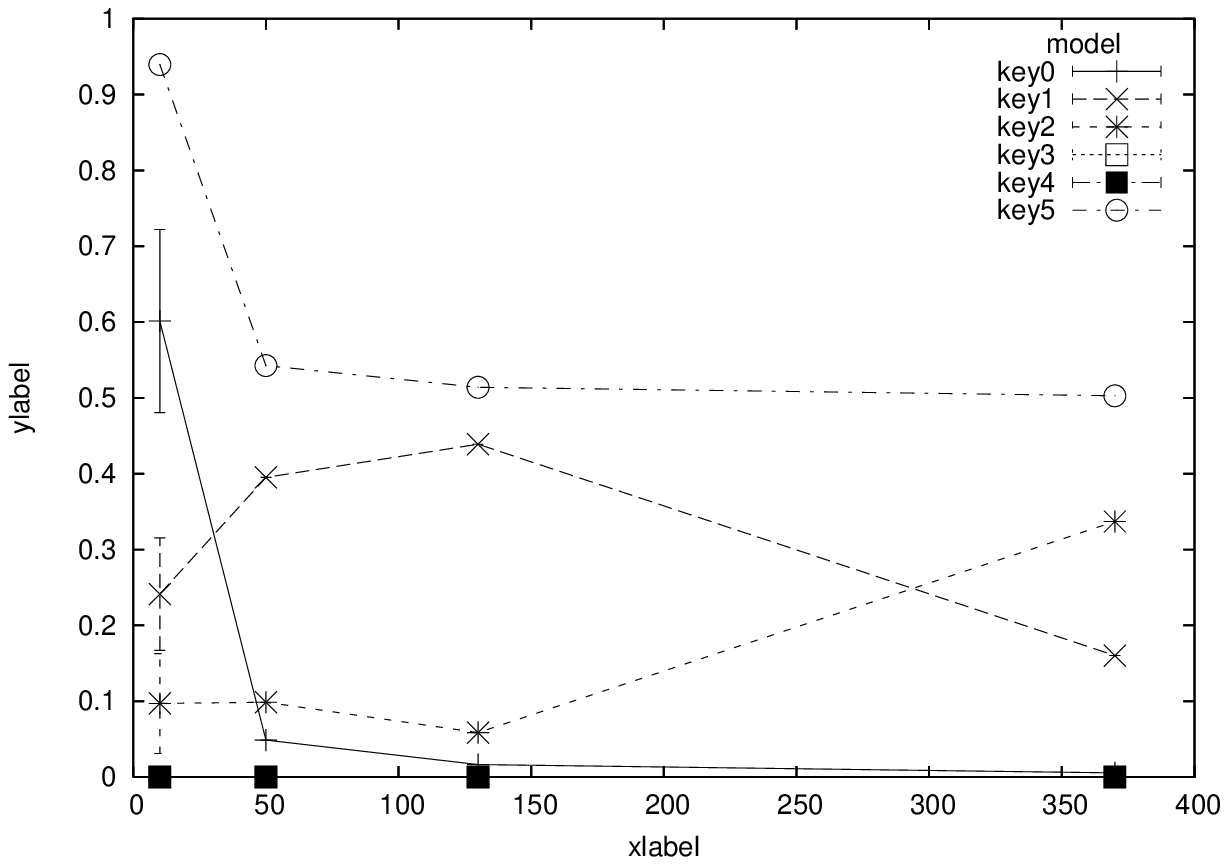}
\caption{\label{fig-met-vdegree-noscale}The well studied metabolic network, above, has a richer local topology than one can see with the usual clustering coefficient, since $C^{1}_{k}$ decreases quickly ($d=1$) with degree, while deeper measures ($d=2,3$) reveal the structure present even in high degrees. \label{fig-hier-vdegree-noscale}The hierarchical model, shown below, has a much simpler structure and does not reflect the importance of distance $2$ relationships present in the real network.}
\end{figure}

FIG.~\ref{fig-met-vdegree-noscale}(metabolic) shows the clustering profile for the metabolic network. We note that based on the usual coefficient alone ($d=1$), one could suggest that more connected metabolites live in an almost unclustered world.

In contrast, the profile not only shows that more pairs of neighbors are related by the second order of clustering than the first, but makes it clear that, as degree raises, clustering only migrates from distance one and two to distance three and four.

In fact, we verify that the sum of $C^{d}_{k}$ up to $d=5$ is close to $1.0$ for all $k$, which implies that almost every two neighbors of a vertex share with it a cycle of length in the order of the network's average distance.

Now, as can be seen from FIG.~\ref{fig-hier-vdegree-noscale}(model), the growth model shares neither of these properties.

Contrary to the metabolic network, distance $2$ does not play a major role on its local topology. Also, the model's sum of all orders of the clustering profile quickly converges to $\frac{1}{2}$ for vertices of high degree, meaning that close to them it has a very different organization from that of the real network, being more open or tree-like.

So, although these networks are both considered small-world networks, in the sense that they have small average distance while maintaining a significant usual clustering coefficient, our observations with the clustering profile sets them clearly apart: there are networks where almost every vertex experiences, in a generalized sense, the small world phenomena, meaning their neighbors are all closely related, while on other networks this effect is restricted to a subset, out of which nodes have groups of neighbors who, if not for them, would live in distant clusters.

This is not only evidence of very different network-growth processes, but we note that this distinction reflects a strikingly different local topology around the high degree vertices, also called ``hubs'' in the literature, which play a central role in many dynamical processes, notably in disease spreading\cite{epidemicstructured-eguiluz} and information retrieval\cite{searchpowerlaw-adamic}. Therefore, establishing this distinction is an important step towards a more structured understanding of both the growth of, and the dynamics on, small-world networks.

From the arguments above, it is useful to define ``complete-small-world networks'' as those networks with small average distance and, for all degrees, the sum of their clustering profile up to an order of that average distance close to 1. As we noted for the metabolic network, this is equivalent to requiring that vertices share short cycles with most of their pairs of neighbors, regardless of degree.

\subsection{Hierarchy}

\begin{figure}
\psfrag{xlabel}[Bc][Bc]{$k$}
\psfrag{ylabel}[Bc][Bc]{$C^{d}_{k}$}
\psfrag{key0}[Br][Br][0.7]{$d\ 1$}
\psfrag{key1}[Br][Br][0.7]{$2$}
\psfrag{key2}[Br][Br][0.7]{$3$}
\psfrag{key3}[Br][Br][0.7]{$4$}
\psfrag{key4}[Br][Br][0.7]{$5$}
\includegraphics[width=\columnwidth]{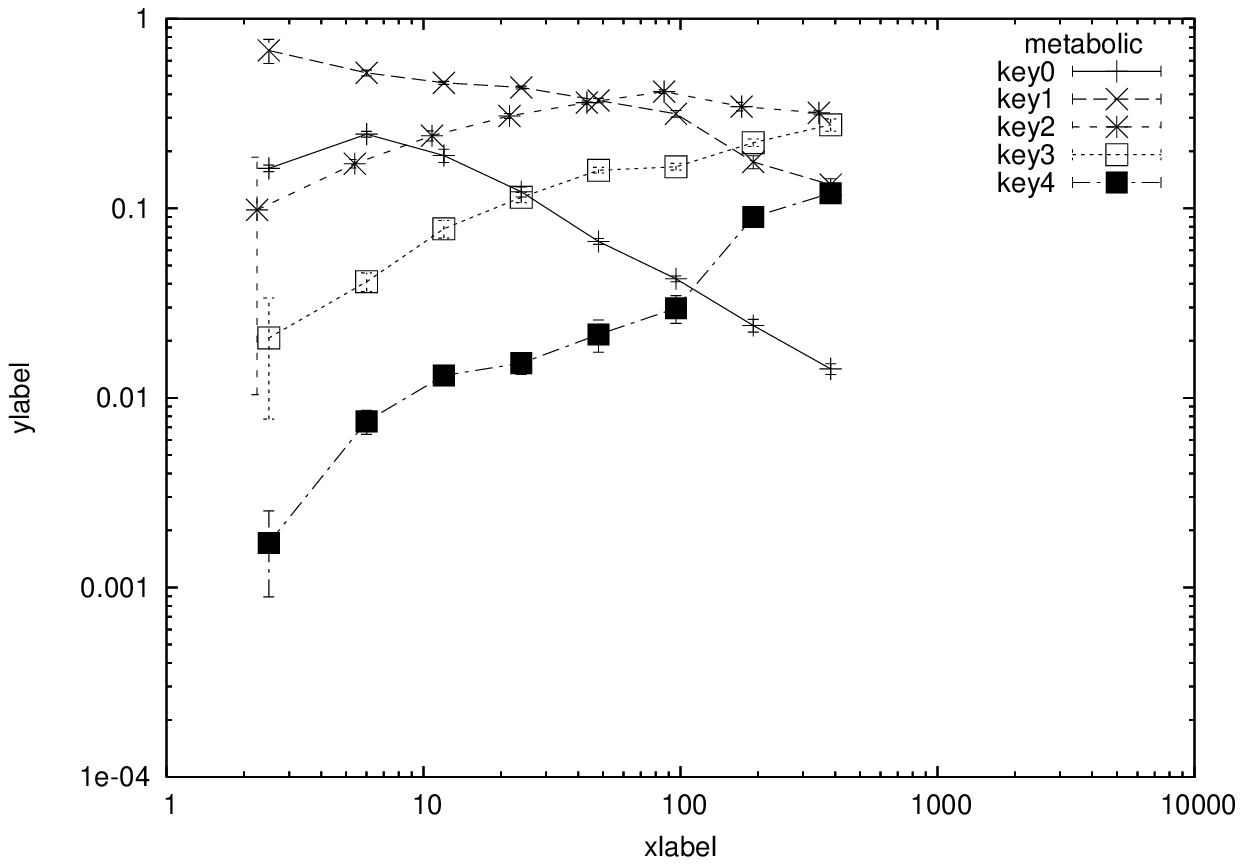}
\includegraphics[width=\columnwidth]{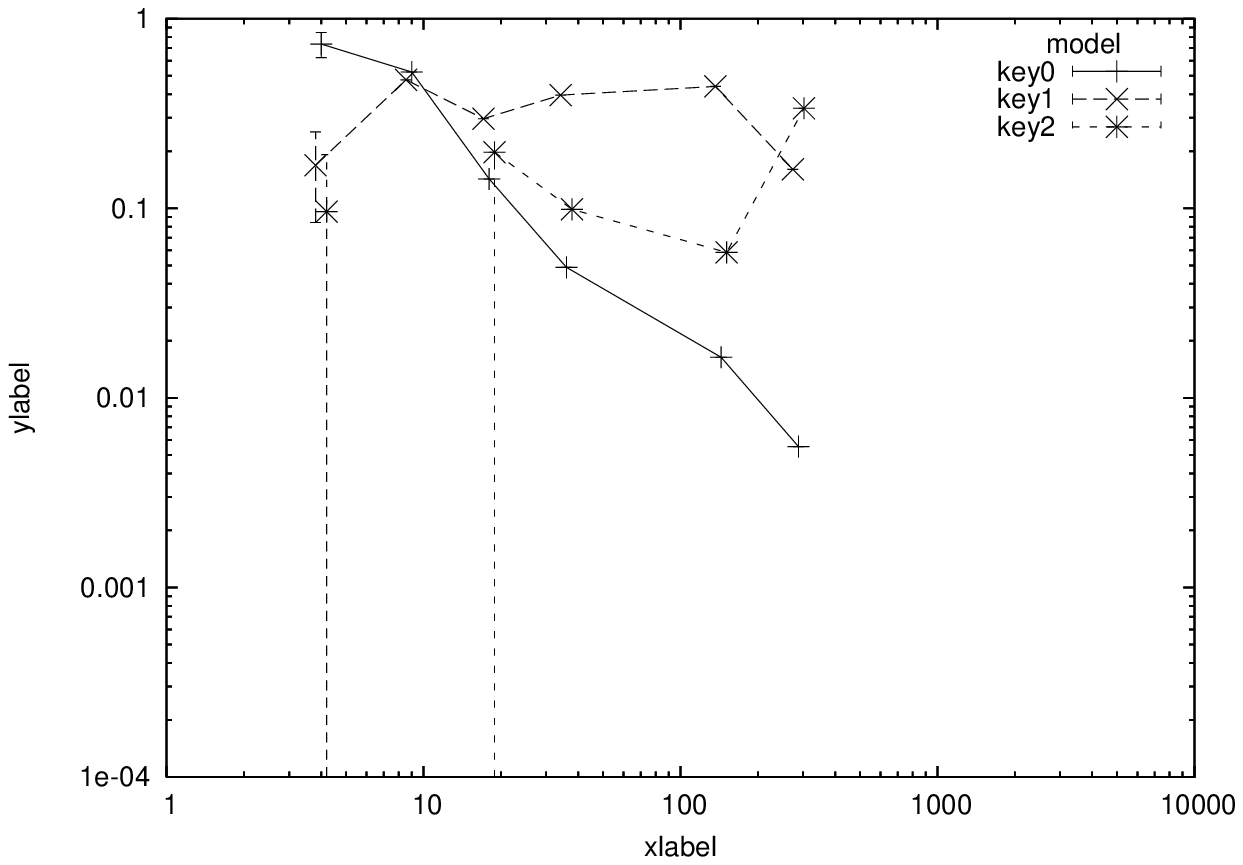}
\caption{\label{fig-met-vdegree-dilog}In a log-log scale, we can see the metabolic network has a wide range with power-law behavior even for higher distances. \label{fig-hier-vdegree-dilog}The model, however, lacks this deep hierarchical structure, and has a zero profile for orders above $3$.}
\end{figure}

Fig.\ref{fig-met-vdegree-dilog} shows the same data scaled to a log-log scale, in order to visualize the power-law behavior of clustering which characterizes hierarchy. We note that, on the metabolic network, $C^{d}_{k}$ for all $d$ varies as a power law in $k$ over a wide range of degrees, indicating a deeper hierarchical structure than previously known.

For the model, however, other than the usual clustering coefficient $C^{1}_{k}$, all additional orders differ from the behavior of the real network: $C^{2}_{k}$ and $C^{3}_{k}$ can hardly be considered hierarchical, and all orders greater than $3$ are constantly equal to zero.

While clearly this model was crafted only to illustrate the idea of hierarchy, we have been able to spot another important feature of the metabolic network's topology, its deep hierarchical structure, that is missing from the model and would be relevant when studying, for example, flow-dynamics\cite{jamming-zoltan,gradientnetworks-zoltan} in bacteria metabolism.

\subsection{Clusterization dynamics}

At last, we consider the behavior of the profile as a function of distance, for specific ranges of degrees, in order to examine the change in influence of growth dynamics over varying orders of clustering. We choose not to use the hierarchical model this time because, its profile being zero beyond order $3$, it is of questionable significance to consider its variation.

Instead, we compare the metabolic network with another small-world network of similar global characteristics, the World Wide Web\cite{diameterweb-barabasi}. Both are scale-free dissortative\cite{mixing-newman} hierarchical\cite{hierarchical-ravasz} networks. Our purpose here is to ask if the evolution from one order to another might give more clues about their underlying clusterization dynamics, since these networks are similar in every other aspect.

However, given that the absolute value of clustering depends on degree, there is little reason to suppose its variation with order would remain the same. Therefore, we split each network into three fractions: small degrees, medium degrees and large degrees, based on the behavior of the degree variation for the clustering profile on these networks. Medium degrees are those where it behaves well as a power law for various orders, small and high degrees are those under and above that range, respectively.

\begin{figure}
\psfrag{xlabel1}[Bc][Bc]{$d$}
\psfrag{ylabel1}[Bc][Bc]{$<C^{d}(u)>$}
\psfrag{xlabel2}[Bc][Bc]{$d^2$}
\psfrag{ylabel2}[Bc][Bc]{$<C^{d}(u)>$}
\includegraphics[width=\columnwidth]{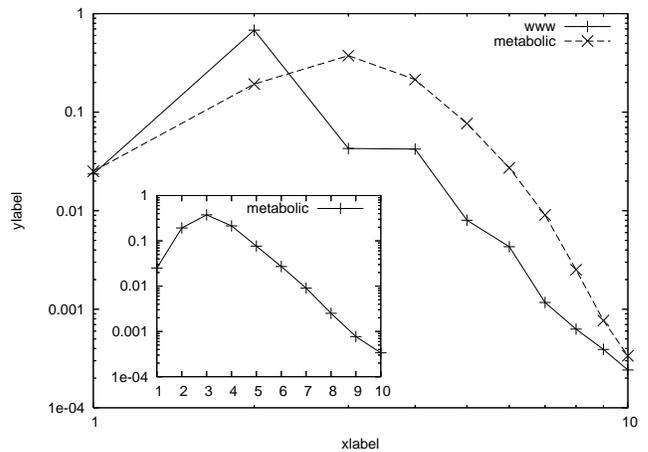}
\caption{\label{fig-vdistance}Variation with distance, in log-log scale, of the average clustering over vertices with high degree. The metabolic network shows exponential behavior (see inset in monolog scale), while the WWW appears as a power law.}
\end{figure}

We verify that, for low and medium degrees, these networks show the same behavior, namely an exponential decrease of the coefficients with increasing order. But as we see on FIG.~\ref{fig-vdistance}, for high degrees the metabolic network remains exponential, while the WWW behaves as a power law, revealing a structural change around its most connected vertices.

This adds a new approach for evaluating and improving various network growth models that use community structure\cite{highlyclustered-newman}, edge addition bias\cite{triadclustering-holme,growingsocial-jin} and other strategies to explain the diverse mechanisms of clustering formation found in natural, social and technological networks. In the present case, it suggests a mechanism of clusterization dynamics more sensible to degree variation for the WWW.

We also observe that the first one or two orders of clustering on these networks do not uniformly follow such scales. In the metabolic network the first order is always lower than extrapolation would suggest and, for high degrees, so is the second order. In the WWW, however, that only happens on high degrees, and only to the first order of clustering, as can be seen in the picture.

It is not clear why such a deviation occurs, but we suppose it is a consequence of superimposed dynamic rules affecting clusterization.

On the metabolic network, we suggest this effect might be related to selective pressure against congestion\cite{jamming-zoltan,jamming-park} of metabolic pathways, since it is always present and affecting deeper levels of clustering, with the effect getting stronger towards highly connected metabolites.

As for the WWW, this poses one interesting question: whether the competing rule is only suppressing the lower orders of the profile on high degrees, or whether its interaction with the dynamics is also the cause of the profile's deviation from exponential behavior.

We intend to pursue these questions on a future study of network growth models.

\section{Conclusions}

In this paper we reviewed some issues, limitations, and alternatives to the clustering coefficient, a measure which is a pivot concept in the field of complex networks. Through a more comprehensive formulation of this concept, new insights were given for networks well studied in the literature, specially when considering network growth processes. Not limited to that, the new measure presents us a broader view of usual phenomena related to network structure dynamics, such as the emergence of clustering, hierarchy, and the small-world effect. It also seems to be most consequential for well connected vertices, which are central in many networked processes. As a final statement, we believe that less specialized, richer descriptions will allow local topology to play a more significant role in understanding the interaction between structure and dynamics on networks.

\section{Material and methods}

The data used in this paper for the metabolic network and the WWW was that available on the website of the CCNR at Univ. of Notre Dame, \url{http://www.nd.edu/~networks/} . The metabolic network was reduced according to the procedure in the supplemental online material of reference \cite{hierarchical-barabasi}.

Non scaled graphics for the profile (Fig.~\ref{fig-met-vdegree-noscale}) were rebinned, log-log graphics for the profile (Fig.~\ref{fig-met-vdegree-dilog}) were log-rebinned, and the error bars are those from the rebinning process.

Higher order clustering coefficients and the clustering profile were calculated with use of the \emph{graph-tool}, which is publicly available as \emph{Free Software}\footnote{Under the terms of the GNU General Public License} at \url{http://graph-tool.forked.de/} .

\begin{acknowledgments}
The authors thank Tiago Peixoto and Al Scandar Solstag for useful conversations and advice. This work was funded in part by Funda\c{c}\~ao de Amparo \`a Pesquisa do Estado de S\~ao Paulo (FAPESP, process 2005/00933-0) and CNPq.
\end{acknowledgments}

\bibliography{abdo}

\begin{thebibliography}{30}
\expandafter\ifx\csname natexlab\endcsname\relax\def\natexlab#1{#1}\fi
\expandafter\ifx\csname bibnamefont\endcsname\relax
  \def\bibnamefont#1{#1}\fi
\expandafter\ifx\csname bibfnamefont\endcsname\relax
  \def\bibfnamefont#1{#1}\fi
\expandafter\ifx\csname citenamefont\endcsname\relax
  \def\citenamefont#1{#1}\fi
\expandafter\ifx\csname url\endcsname\relax
  \def\url#1{\texttt{#1}}\fi
\expandafter\ifx\csname urlprefix\endcsname\relax\def\urlprefix{URL }\fi
\providecommand{\bibinfo}[2]{#2}
\providecommand{\eprint}[2][]{\url{#2}}

\bibitem[{\citenamefont{S.~Bornholdt}(2003)}]{graphshandbook-editors}
\bibinfo{editor}{\bibfnamefont{H.~G.~S.} \bibnamefont{S.~Bornholdt}}, ed.,
  \emph{\bibinfo{title}{Handbook of Graphs and Networks: From the Genome to the
  Internet}} (\bibinfo{publisher}{Wiley-VCH}, \bibinfo{address}{Berlin},
  \bibinfo{year}{2003}).

\bibitem[{\citenamefont{Dorogovtsev and
  Mendes}(2003)}]{evolutionnetworks-dorogovtsev}
\bibinfo{author}{\bibfnamefont{S.~N.} \bibnamefont{Dorogovtsev}}
  \bibnamefont{and} \bibinfo{author}{\bibfnamefont{J.~F.~F.}
  \bibnamefont{Mendes}}, \emph{\bibinfo{title}{Evolution of Networks: From
  Biological Nets to the Internet and WWW}} (\bibinfo{publisher}{Oxford
  University Press}, \bibinfo{address}{USA}, \bibinfo{year}{2003}).

\bibitem[{\citenamefont{Diestel}(1997)}]{graphs-diestel}
\bibinfo{author}{\bibfnamefont{R.}~\bibnamefont{Diestel}},
  \emph{\bibinfo{title}{Graph Theory}} (\bibinfo{publisher}{Springer-Verlag,
  Heidelberg}, \bibinfo{address}{New York}, \bibinfo{year}{1997}).

\bibitem[{\citenamefont{Newman}(2003{\natexlab{a}})}]{networksreview-newman}
\bibinfo{author}{\bibfnamefont{M.~E.~J.} \bibnamefont{Newman}},
  \bibinfo{journal}{SIAM Review} \textbf{\bibinfo{volume}{45}},
  \bibinfo{pages}{167} (\bibinfo{year}{2003}{\natexlab{a}}).

\bibitem[{\citenamefont{Bianconi and Capocci}(2003)}]{loopiness-bianconi}
\bibinfo{author}{\bibfnamefont{G.}~\bibnamefont{Bianconi}} \bibnamefont{and}
  \bibinfo{author}{\bibfnamefont{A.}~\bibnamefont{Capocci}},
  \bibinfo{journal}{Physical Review Letters} \textbf{\bibinfo{volume}{90}},
  \bibinfo{pages}{078701} (\bibinfo{year}{2003}).

\bibitem[{\citenamefont{Milo et~al.}(2002)\citenamefont{Milo, Shen-Orr,
  Itzkovitz, Kashtan, Chklovskii, and Alon}}]{motifs-milo}
\bibinfo{author}{\bibfnamefont{R.}~\bibnamefont{Milo}},
  \bibinfo{author}{\bibfnamefont{S.}~\bibnamefont{Shen-Orr}},
  \bibinfo{author}{\bibfnamefont{S.}~\bibnamefont{Itzkovitz}},
  \bibinfo{author}{\bibfnamefont{N.}~\bibnamefont{Kashtan}},
  \bibinfo{author}{\bibfnamefont{D.}~\bibnamefont{Chklovskii}},
  \bibnamefont{and} \bibinfo{author}{\bibfnamefont{U.}~\bibnamefont{Alon}},
  \bibinfo{journal}{Science} \textbf{\bibinfo{volume}{298}},
  \bibinfo{pages}{824} (\bibinfo{year}{2002}).

\bibitem[{\citenamefont{Watts and Strogatz}(1998)}]{smallworld-watts}
\bibinfo{author}{\bibfnamefont{D.~J.} \bibnamefont{Watts}} \bibnamefont{and}
  \bibinfo{author}{\bibfnamefont{S.~H.} \bibnamefont{Strogatz}},
  \bibinfo{journal}{Nature} \textbf{\bibinfo{volume}{393}},
  \bibinfo{pages}{440} (\bibinfo{year}{1998}).

\bibitem[{\citenamefont{Zhang et~al.}(2002)\citenamefont{Zhang, Goel, and
  Govindan}}]{freenet-zhang}
\bibinfo{author}{\bibfnamefont{H.}~\bibnamefont{Zhang}},
  \bibinfo{author}{\bibfnamefont{A.}~\bibnamefont{Goel}}, \bibnamefont{and}
  \bibinfo{author}{\bibfnamefont{R.}~\bibnamefont{Govindan}},
  \bibinfo{journal}{IEEE INFOCOM} \textbf{\bibinfo{volume}{3}},
  \bibinfo{pages}{1228} (\bibinfo{year}{2002}).

\bibitem[{\citenamefont{Watts et~al.}(2002)\citenamefont{Watts, Dodds, and
  Newman}}]{identitysearch-watts}
\bibinfo{author}{\bibfnamefont{D.~J.} \bibnamefont{Watts}},
  \bibinfo{author}{\bibfnamefont{P.~S.} \bibnamefont{Dodds}}, \bibnamefont{and}
  \bibinfo{author}{\bibfnamefont{M.~E.~J.} \bibnamefont{Newman}},
  \bibinfo{journal}{Science} \textbf{\bibinfo{volume}{296}},
  \bibinfo{pages}{1302} (\bibinfo{year}{2002}).

\bibitem[{\citenamefont{Cross et~al.}(2004)\citenamefont{Cross, Lloyd-Smith,
  Bowers, Hay, Hofmeyr, and Getz}}]{clusteringdisease-cross}
\bibinfo{author}{\bibfnamefont{P.~C.} \bibnamefont{Cross}},
  \bibinfo{author}{\bibfnamefont{J.~O.} \bibnamefont{Lloyd-Smith}},
  \bibinfo{author}{\bibfnamefont{J.~A.} \bibnamefont{Bowers}},
  \bibinfo{author}{\bibfnamefont{C.~T.} \bibnamefont{Hay}},
  \bibinfo{author}{\bibfnamefont{M.}~\bibnamefont{Hofmeyr}}, \bibnamefont{and}
  \bibinfo{author}{\bibfnamefont{W.~M.} \bibnamefont{Getz}},
  \bibinfo{journal}{Ann. Zool. Fennici} \textbf{\bibinfo{volume}{41}},
  \bibinfo{pages}{879} (\bibinfo{year}{2004}).

\bibitem[{\citenamefont{Egu\'{i}luz and
  Klemm}(2002)}]{epidemicstructured-eguiluz}
\bibinfo{author}{\bibfnamefont{V.~M.} \bibnamefont{Egu\'{i}luz}}
  \bibnamefont{and} \bibinfo{author}{\bibfnamefont{K.}~\bibnamefont{Klemm}},
  \bibinfo{journal}{Physical Review Letters} \textbf{\bibinfo{volume}{89}},
  \bibinfo{pages}{108701} (\bibinfo{year}{2002}).

\bibitem[{\citenamefont{Serrano and
  Bogun\'a}(2006)}]{serrano-generalclustering}
\bibinfo{author}{\bibfnamefont{M.~A.} \bibnamefont{Serrano}} \bibnamefont{and}
  \bibinfo{author}{\bibfnamefont{M.}~\bibnamefont{Bogun\'a}}
  (\bibinfo{year}{2006}), \eprint{cond-mat/0608336}.

\bibitem[{\citenamefont{Newman et~al.}(2001)\citenamefont{Newman, Strogatz, and
  Watts}}]{generatingfunctions-newman}
\bibinfo{author}{\bibfnamefont{M.~E.~J.} \bibnamefont{Newman}},
  \bibinfo{author}{\bibfnamefont{S.~H.} \bibnamefont{Strogatz}},
  \bibnamefont{and} \bibinfo{author}{\bibfnamefont{D.~J.} \bibnamefont{Watts}},
  \bibinfo{journal}{Physical Review E} \textbf{\bibinfo{volume}{64}},
  \bibinfo{pages}{026118} (\bibinfo{year}{2001}).

\bibitem[{\citenamefont{Newman and Park}(2003)}]{socialnetworks-newman}
\bibinfo{author}{\bibfnamefont{M.~E.~J.} \bibnamefont{Newman}}
  \bibnamefont{and} \bibinfo{author}{\bibfnamefont{J.}~\bibnamefont{Park}},
  \bibinfo{journal}{Physical Review E} \textbf{\bibinfo{volume}{68}},
  \bibinfo{pages}{036122} (\bibinfo{year}{2003}).

\bibitem[{\citenamefont{Bollob\'{a}s and Riordan}(2003)}]{handbook-bollobas}
\bibinfo{author}{\bibfnamefont{B.}~\bibnamefont{Bollob\'{a}s}}
  \bibnamefont{and} \bibinfo{author}{\bibfnamefont{O.~M.}
  \bibnamefont{Riordan}}, \emph{\bibinfo{title}{Handbook of Graphs and
  Networks: From the Genome to the Internet}} (\bibinfo{publisher}{Wiley-VCH},
  \bibinfo{address}{Weinheim}, \bibinfo{year}{2003}), \bibinfo{note}{pp. 1-34}.

\bibitem[{\citenamefont{Ravasz et~al.}(2002)\citenamefont{Ravasz, Somera,
  Mongru, Oltvai, and Barab\'{a}si}}]{hierarchical-barabasi}
\bibinfo{author}{\bibfnamefont{E.}~\bibnamefont{Ravasz}},
  \bibinfo{author}{\bibfnamefont{A.~L.} \bibnamefont{Somera}},
  \bibinfo{author}{\bibfnamefont{D.~A.} \bibnamefont{Mongru}},
  \bibinfo{author}{\bibfnamefont{Z.~N.} \bibnamefont{Oltvai}},
  \bibnamefont{and} \bibinfo{author}{\bibfnamefont{A.-L.}
  \bibnamefont{Barab\'{a}si}}, \bibinfo{journal}{Science}
  \textbf{\bibinfo{volume}{297}}, \bibinfo{pages}{1551} (\bibinfo{year}{2002}).

\bibitem[{\citenamefont{Lind et~al.}(2005)\citenamefont{Lind, Gonz\'{a}lez, and
  Herrmann}}]{bipartite-herrmann}
\bibinfo{author}{\bibfnamefont{P.~G.} \bibnamefont{Lind}},
  \bibinfo{author}{\bibfnamefont{M.~C.} \bibnamefont{Gonz\'{a}lez}},
  \bibnamefont{and} \bibinfo{author}{\bibfnamefont{H.~J.}
  \bibnamefont{Herrmann}}, \bibinfo{journal}{Physical Review E}
  \textbf{\bibinfo{volume}{72}}, \bibinfo{pages}{056127}
  (\bibinfo{year}{2005}).

\bibitem[{\citenamefont{Fronczak et~al.}(2002)\citenamefont{Fronczak, Holyst,
  Jedynak, and Sienkiewicz}}]{higherorder-fronczak}
\bibinfo{author}{\bibfnamefont{A.}~\bibnamefont{Fronczak}},
  \bibinfo{author}{\bibfnamefont{J.~A.} \bibnamefont{Holyst}},
  \bibinfo{author}{\bibfnamefont{M.}~\bibnamefont{Jedynak}}, \bibnamefont{and}
  \bibinfo{author}{\bibfnamefont{J.}~\bibnamefont{Sienkiewicz}},
  \bibinfo{journal}{Physica A} \textbf{\bibinfo{volume}{316}},
  \bibinfo{pages}{688} (\bibinfo{year}{2002}).

\bibitem[{\citenamefont{Barab\'{a}si and Albert}(1999)}]{scalefree-barabasi}
\bibinfo{author}{\bibfnamefont{A.-L.} \bibnamefont{Barab\'{a}si}}
  \bibnamefont{and} \bibinfo{author}{\bibfnamefont{R.}~\bibnamefont{Albert}},
  \bibinfo{journal}{Science} \textbf{\bibinfo{volume}{286}},
  \bibinfo{pages}{509} (\bibinfo{year}{1999}).

\bibitem[{\citenamefont{Jeong et~al.}(2000)\citenamefont{Jeong, Tombor, Albert,
  Oltvai, and Barab\'{a}si}}]{largescalemetabolic-barabasi}
\bibinfo{author}{\bibfnamefont{H.}~\bibnamefont{Jeong}},
  \bibinfo{author}{\bibfnamefont{B.}~\bibnamefont{Tombor}},
  \bibinfo{author}{\bibfnamefont{R.}~\bibnamefont{Albert}},
  \bibinfo{author}{\bibfnamefont{Z.~N.} \bibnamefont{Oltvai}},
  \bibnamefont{and} \bibinfo{author}{\bibfnamefont{A.-L.}
  \bibnamefont{Barab\'{a}si}}, \bibinfo{journal}{Nature}
  \textbf{\bibinfo{volume}{407}}, \bibinfo{pages}{651} (\bibinfo{year}{2000}).

\bibitem[{\citenamefont{Adamic et~al.}(2001)\citenamefont{Adamic, Lukose,
  Puniyani, and Huberman}}]{searchpowerlaw-adamic}
\bibinfo{author}{\bibfnamefont{L.~A.} \bibnamefont{Adamic}},
  \bibinfo{author}{\bibfnamefont{R.~M.} \bibnamefont{Lukose}},
  \bibinfo{author}{\bibfnamefont{A.~R.} \bibnamefont{Puniyani}},
  \bibnamefont{and} \bibinfo{author}{\bibfnamefont{B.~A.}
  \bibnamefont{Huberman}}, \bibinfo{journal}{Physical Review E}
  \textbf{\bibinfo{volume}{64}}, \bibinfo{pages}{046135}
  (\bibinfo{year}{2001}).

\bibitem[{\citenamefont{Toroczkai and Bassler}(2004)}]{jamming-zoltan}
\bibinfo{author}{\bibfnamefont{Z.}~\bibnamefont{Toroczkai}} \bibnamefont{and}
  \bibinfo{author}{\bibfnamefont{K.~E.} \bibnamefont{Bassler}},
  \bibinfo{journal}{Nature} \textbf{\bibinfo{volume}{428}},
  \bibinfo{pages}{716} (\bibinfo{year}{2004}).

\bibitem[{\citenamefont{Toroczkai et~al.}(2004)\citenamefont{Toroczkai, Kozma,
  Bassler, Hengartner, and Korniss}}]{gradientnetworks-zoltan}
\bibinfo{author}{\bibfnamefont{Z.}~\bibnamefont{Toroczkai}},
  \bibinfo{author}{\bibfnamefont{B.}~\bibnamefont{Kozma}},
  \bibinfo{author}{\bibfnamefont{K.~E.} \bibnamefont{Bassler}},
  \bibinfo{author}{\bibfnamefont{N.}~\bibnamefont{Hengartner}},
  \bibnamefont{and} \bibinfo{author}{\bibfnamefont{G.}~\bibnamefont{Korniss}}
  (\bibinfo{year}{2004}), \eprint{cond-mat/0408262}.

\bibitem[{\citenamefont{R.~Albert and
  Barab\'{a}si}(1999)}]{diameterweb-barabasi}
\bibinfo{author}{\bibfnamefont{H.~J.} \bibnamefont{R.~Albert}}
  \bibnamefont{and} \bibinfo{author}{\bibfnamefont{A.-L.}
  \bibnamefont{Barab\'{a}si}}, \bibinfo{journal}{Nature}
  \textbf{\bibinfo{volume}{401}}, \bibinfo{pages}{130} (\bibinfo{year}{1999}).

\bibitem[{\citenamefont{Newman}(2003{\natexlab{b}})}]{mixing-newman}
\bibinfo{author}{\bibfnamefont{M.~E.~J.} \bibnamefont{Newman}},
  \bibinfo{journal}{Physical Review E} \textbf{\bibinfo{volume}{67}},
  \bibinfo{pages}{026126} (\bibinfo{year}{2003}{\natexlab{b}}).

\bibitem[{\citenamefont{Ravasz and Barab\'{a}si}(2003)}]{hierarchical-ravasz}
\bibinfo{author}{\bibfnamefont{E.}~\bibnamefont{Ravasz}} \bibnamefont{and}
  \bibinfo{author}{\bibfnamefont{A.-L.} \bibnamefont{Barab\'{a}si}},
  \bibinfo{journal}{Physical Review E} \textbf{\bibinfo{volume}{67}},
  \bibinfo{pages}{026112} (\bibinfo{year}{2003}).

\bibitem[{\citenamefont{Newman}(2003{\natexlab{c}})}]{highlyclustered-newman}
\bibinfo{author}{\bibfnamefont{M.~E.~J.} \bibnamefont{Newman}},
  \bibinfo{journal}{Physical Review E} \textbf{\bibinfo{volume}{68}},
  \bibinfo{pages}{026121} (\bibinfo{year}{2003}{\natexlab{c}}).

\bibitem[{\citenamefont{Holme and Kim}(2002)}]{triadclustering-holme}
\bibinfo{author}{\bibfnamefont{P.}~\bibnamefont{Holme}} \bibnamefont{and}
  \bibinfo{author}{\bibfnamefont{B.~J.} \bibnamefont{Kim}},
  \bibinfo{journal}{Physical Review E} \textbf{\bibinfo{volume}{65}},
  \bibinfo{pages}{026107} (\bibinfo{year}{2002}).

\bibitem[{\citenamefont{Jin et~al.}(2001)\citenamefont{Jin, Girvan, and
  Newman}}]{growingsocial-jin}
\bibinfo{author}{\bibfnamefont{E.~M.} \bibnamefont{Jin}},
  \bibinfo{author}{\bibfnamefont{M.}~\bibnamefont{Girvan}}, \bibnamefont{and}
  \bibinfo{author}{\bibfnamefont{M.~E.~J.} \bibnamefont{Newman}},
  \bibinfo{journal}{Physical Review E} \textbf{\bibinfo{volume}{64}},
  \bibinfo{pages}{046132} (\bibinfo{year}{2001}).

\bibitem[{\citenamefont{Park et~al.}(2005)\citenamefont{Park, Lai, Zhao, and
  Ye}}]{jamming-park}
\bibinfo{author}{\bibfnamefont{K.}~\bibnamefont{Park}},
  \bibinfo{author}{\bibfnamefont{Y.-C.} \bibnamefont{Lai}},
  \bibinfo{author}{\bibfnamefont{L.}~\bibnamefont{Zhao}}, \bibnamefont{and}
  \bibinfo{author}{\bibfnamefont{N.}~\bibnamefont{Ye}},
  \bibinfo{journal}{Physical Review E} \textbf{\bibinfo{volume}{71}},
  \bibinfo{pages}{065105(R)} (\bibinfo{year}{2005}).

\end{thebibliography}

\end{document}